# Simple derivation of the frequency dependent complex heat capacity

J.-L. Garden*

*Institut Néel, CNRS et Université Joseph Fourier, BP 166, 38042 Grenoble Cedex 9, France.*

**Abstract**

*This paper gives a simple derivation of the well-known expression of the frequency dependent complex heat capacity in modulated temperature experiments. It aims at clarified again that the generalized calorimetric susceptibility is only due to the non-equilibrium behaviour occurring in the vicinity of thermodynamic equilibrium of slow internal degrees of freedom of a sample when the temperature oscillates at a well determined frequency.*

* Corresponding author. Fax: 33 (0) 4 76 87 50 60

*E-mail address:* jean-luc.garden@grenoble.cnrs.fr (J.-L. Garden)

## 1. Introduction

The frequency dependent complex heat capacity, or also the generalized calorimetric susceptibility, can be encountered in the literature under this following well-known form:

$$C^* = C' - iC'' = C_\infty + \frac{C_0 - C_\infty}{1 + i\omega\tau} \quad (1)$$

where

$$C' = C_\infty + \frac{C_0 - C_\infty}{1 + (\omega\tau)^2} \quad (2)$$

is the storage frequency dependent heat capacity, and:

$$C'' = \frac{(C_0 - C_\infty)\omega\tau}{1 + (\omega\tau)^2} \quad (3)$$

is the loss frequency dependent heat capacity [1]. $C_\infty$ is the heat capacity related to the infinitely fast degrees of freedom of the system as compared to the frequency (generally vibrational modes or phonons bath), and $C_0$ is the total contribution at equilibrium (the frequency is set to zero) of the degrees of freedom, fast and slow, of the sample. The time constant $\tau$ is the kinetic relaxation time constant of a certain internal degree of freedom. In the following, we assume that these equations are well known, and the reader interested by a deeper insight on this subject can find more details in the following references list [2-11]. In this short communication, we show that these last formulas can be derived in a very simple way using only few basic assumptions.

**2. Heat capacity at thermodynamic equilibrium**

At constant pressure, and when no work is involved in a thermodynamic reversible transformation, the first law of thermodynamics can be enunciated as follows:

$$\delta Q = dH \quad (4)$$

where $\delta Q$ is the quantity of heat exchanged reversibly between a system and the surroundings, and $H$ is the state function enthalpy of the system.

Let us focus on a particular internal degree of freedom inside a thermodynamic system. For simplicity, we can observe the evolvement of a simple chemical reaction. In this case, the internal degree of freedom is the evolution of the reaction and it is characterized by a parameter wich is often called, $\xi$, the degree of advance of the chemical reaction or extent of the reaction. Nevertheless, we can choose any others internal degree of freedom such as vibrational modes of the internal structure of a molecule, translational or rotational modes within the system, or by an order parameter characterizing the advance of a phase transition. For simplicity, by now $\xi$ is called the order parameter of the internal degree of freedom. Hence, the thermodynamic system can now be characterized by two thermodynamic variables which are the temperature $T$ and the order parameter $\xi$. The enthalpy of the system can be simply differentiated with respect to these two variables:

$$dH = \partial H/\partial T)_\xi \, dT + \partial H/\partial \xi)_T \, d\xi \quad (5)$$

The heat capacity is defined as the ratio of the quantity of heat exchanged between the system and the surroundings with the temperature variation recorded by the experimentalist:

$$C_{mes} = \frac{dH}{dT} = \partial H/\partial T)_\xi + \partial H/\partial \xi)_T \frac{d\xi}{dT} \quad (6)$$

If the transformation is a reversible thermodynamic transformation, the system is at any time in a state of equilibrium. This is to say that the couple of thermodynamic variables *(T, ξ)* have always their equilibrium values *($T_{eq}$, $\xi_{eq}$)* at any time. For the temperature, this must say that firstly the temperature is homogenous in all part of the system at any time (infinite thermal diffusivity) and secondly that the temperature has had no time to relax towards the thermal bath during the transformation (calorimetric adiabaticity conditions). For the variable *ξ*, this simply means that the degree of freedom has a kinetic relaxation time fast enough to instantaneously follow the temperature variation over the time interval of the transformation:

$$\xi_{eq}(t) = \xi_{eq}[T_{eq}(t)] \quad (7)$$

From (6) the temperature derivative of the order parameter appearing in the right-hand side can be expressed as:

$$\frac{d\xi_{eq}}{dT} = \frac{[C_{mes}^{eq} - \partial H/\partial T)_\xi^{eq}]}{\partial H/\partial \xi)_T^{eq}} \quad (8)$$

Let keep this expression in our memory, but we can remark already here that, if the measured heat capacity is inferred from a temperature modulated calorimetric experiment, then at equilibrium it can be identified with $C_0$ of the formula (1). Secondly, the temperature derivative of the enthalpy at constant order parameter can be identified with $C_\infty$. Indeed, this latter term is just equivalent to the heat capacity at constant composition of the system, as if the order parameter was frozen during the temperature variation over the time scale of the experiment. Thus, for a reversible transformation we have:

$$\frac{d\xi_{eq}}{dT} = \frac{[C_0 - C_\infty]}{\partial H / \partial \xi)_T^{eq}} \quad (9)$$

## 3. Linear regime close to thermodynamic equilibrium

Now, imagine that under the variation $\Delta T$, over the finite time scale $\Delta t$, the kinetic relaxation time constant $\tau$ of the considered internal degree of freedom is so high that its order parameter has had not enough time to relax towards its thermodynamic equilibrium value $\xi_{eq}$ under the finite temporal variation $\Delta t$. In this case, the system is out of its thermodynamic equilibrium state with respect to the variable $\xi$, and this variable varies along the time over the time scale of the measurement. Now, the most important hypothesis consists to assume that, if the temperature increment is small enough then the system keeps in the vicinity of equilibrium. This condition can mathematically be expressed as follows:

$$\frac{d\xi}{dt} = -\alpha[\xi(t) - \xi_{eq}(t)] \quad (10)$$

This is to say, the order parameter rate in the vicinity of equilibrium is just proportional to the departure of the variable $\xi$ from its value at equilibrium. When it is brought outside equilibrium $\xi$ relaxes exponentially towards its equilibrium value. Without justification, it is interesting to indicate that the proportionality coefficient $\alpha$ is positive for the stability of the equilibrium state and also that it is identical to the inverse of the kinetic relaxation time $\tau$ (see also [6]).

**4. Temperature oscillation**

The first order linear differential equation (10) can be more explicitly rewritten:

$$\tau \delta\dot{\xi} + \delta\xi = \delta\xi_{eq} \quad (11)$$

where $\begin{cases} \delta\xi = \xi(t) - \xi_{eq}^{dc} \\ \delta\xi_{eq}(t) = \xi_{eq}(t) - \xi_{eq}^{dc} \end{cases} \quad (12)$

The dot represents the time derivative, and $\xi_{eq}^{dc}$ is the constant stationary equilibrium value of the variable $\xi$. All the variations of the different parameters are referenced to a constant dc value, which for instance defines the stationary conditions of the experiment

In the case of an harmonic modulation of the temperature, the equilibrium value of the order parameter follows instantaneously the temperature oscillation, and the forcing term of (11) can be written in the linear regime:

$$\delta\xi_{eq}(t) = \frac{d\xi_{eq}}{dT} T_{ac}(t) \quad (13)$$

$T_{ac}(t)$ is the oscillatory term of the temperature around the mean constant temperature $T_{dc}$, which in fact the condition of sationnarity for the temperature ($T_{ac}(t) = T(t) - T_{dc}$)

The resolution of (11) yields simply:

$$\delta\xi(t) = \frac{\frac{d\xi_{eq}}{dT} T_{ac}}{1 + i\omega\tau} \quad (14)$$

which can be explicitly written with the equation (9):

$$\delta\xi(t) = \frac{(C_0 - C_\infty) T_{ac}}{\partial H/\partial \xi)_T^{eq} (1 + i\omega\tau)} \quad (15)$$

**5. Heat capacity out of thermodynamic equilibrium: generalized calorimetric susceptibility**

Even for a small departure from thermodynamic equilibrium the time becomes a preponderant variable. The expression (6) of the measured heat capacity becomes near equilibrium:

$$C_{mes} = C_\infty + \partial H/\partial \xi)_T^{eq} \frac{d\xi/dt}{dT/dt} \quad (16)$$

Including directly the derivative of (15) into (16) gives:

$$C_{mes} = C_\infty + \frac{C_0 - C_\infty}{1 + i\omega\tau} \quad (17)$$

which is the expected expression of the frequency dependent complex heat capacity.

**6. Conclusion**

This paper has recalled on a very simple manner that the frequency dependent complex heat capacity is only due to the departure from equilibrium of an internal degree of freedom of a sample when it is perturbed by an harmonic temperature oscillation. The important assumptions necessary to the validity of the notion of generalized calorimetric susceptibility is not only that the system must be out of its thermodynamic equilibrium state during temperature perturbations, but also more precisely that it must remain in the vicinity of equilibrium in the linear range. In this regime, the time dependent parameters have exponential relaxation towards equilibrium, and the constant parameters take their equilibrium values. Obviously, it is the same assumptions than in the linear response theory approach. Another important point is to consider that variations of the thermodynamic parameters $T$ and $\xi$ are always referenced to a constant quasi-equilibrium state ($T_{dc}, \xi_{eq}^{dc}$) which does not evolve along the time over the time scale of the experiment. This latter point defines the stationary conditions of the experiment.

thank O. Bourgeois for having point out the importance of the stationary conditions in the equation (13).